\documentstyle[psfig]{mn}

\newif\ifAMStwofonts

\newcommand{\ffffff}[1]{\mbox{$#1$}}
\newcommand{\ltsim}{\raisebox{-0.5ex}{$\;\stackrel{<}{\scriptstyle \sim}\;$}}

\newcommand{\simlt}{\ltsim}

\newcommand{\unit}[1]{\ifmmode \:\mbox{\rm #1}\else \mbox{#1}\fi}

\newcommand{\mone}{\ffffff{^{-1}}}

\newcommand{\etal}{{\em et al.\ }}
\newcommand{\eg}{{\em e.g.\/}}

\newcommand{\ie}{{\em i.e.\/}}

\newcommand{\kms}{\unit{km~s\mone}}

\newcommand{\parsec}{\unit{pc}}

\newcommand{\scnd}{\mbox{\ffffff{''}\hskip-0.3em .}}

\def \n1{N$^{+}$} 
\def \o1{O$^+$}

\def\la{\mathrel{\mathchoice {\vcenter{\offinterlineskip\halign{\hfil
$\displaystyle##$\hfil\cr<\cr\sim\cr}}}
{\vcenter{\offinterlineskip\halign{\hfil$\textstyle##$\hfil\cr
<\cr\sim\cr}}}
{\vcenter{\offinterlineskip\halign{\hfil$\scriptstyle##$\hfil\cr
<\cr\sim\cr}}}
{\vcenter{\offinterlineskip\halign{\hfil$\scriptscriptstyle##$\hfil\cr
<\cr\sim\cr}}}}}
\def\ga{\mathrel{\mathchoice {\vcenter{\offinterlineskip\halign{\hfil
$\displaystyle##$\hfil\cr>\cr\sim\cr}}}
{\vcenter{\offinterlineskip\halign{\hfil$\textstyle##$\hfil\cr
>\cr\sim\cr}}}
{\vcenter{\offinterlineskip\halign{\hfil$\scriptstyle##$\hfil\cr
>\cr\sim\cr}}}
{\vcenter{\offinterlineskip\halign{\hfil$\scriptscriptstyle##$\hfil\cr
>\cr\sim\cr}}}}}

\ifoldfss
  \ifCUPmtlplainloaded \else
    \NewTextAlphabet{textbfit} {cmbxti10} {}
    \NewTextAlphabet{textbfss} {cmssbx10} {}
    \NewMathAlphabet{mathbfit} {cmbxti10} {} 
    \NewMathAlphabet{mathbfss} {cmssbx10} {} 
  \fi
  \ifAMStwofonts
    \ifCUPmtlplainloaded \else
      \NewSymbolFont{upmath} {eurm10}
      \NewSymbolFont{AMSa} {msam10}
      \NewMathSymbol{\upi}     {0}{upmath}{19}
      \NewMathSymbol{\umu}     {0}{upmath}{16}
      \NewMathSymbol{\upartial}{0}{upmath}{40}
      \NewMathSymbol{\leqslant}{3}{AMSa}{36}
      \NewMathSymbol{\geqslant}{3}{AMSa}{3E}

    \fi
  \fi
\fi 

\ifnfssone
  \newmathalphabet{\mathit}
  \addtoversion{normal}{\mathit}{cmr}{m}{it}
  \addtoversion{bold}{\mathit}{cmr}{bx}{it}
  \newmathalphabet{\mathbfit} 
  \addtoversion{normal}{\mathbfit}{cmr}{bx}{it}
  \addtoversion{bold}{\mathbfit}{cmr}{bx}{it}
  \newmathalphabet{\mathbfss} 
  \addtoversion{normal}{\mathbfss}{cmss}{bx}{n}
  \addtoversion{bold}{\mathbfss}{cmss}{bx}{n}
  \ifAMStwofonts
    \ifCUPmtlplainloaded \else
      \UseAMStwoboldmath
      \makeatletter
      \new@mathgroup\upmath@group
      \define@mathgroup\mv@normal\upmath@group{eur}{m}{n}
      \define@mathgroup\mv@bold\upmath@group{eur}{b}{n}
      \edef\UPM{\hexnumber\upmath@group}
      \new@mathgroup\amsa@group
      \define@mathgroup\mv@normal\amsa@group{msa}{m}{n}
      \define@mathgroup\mv@bold\amsa@group{msa}{m}{n}
      \edef\AMSa{\hexnumber\amsa@group}
      \makeatother
      \mathchardef\upi="0\UPM19
      \mathchardef\umu="0\UPM16
      \mathchardef\upartial="0\UPM40
      \mathchardef\leqslant="3\AMSa36
      \mathchardef\geqslant="3\AMSa3E
    \fi
  \fi
\fi 

\ifnfsstwo
  \DeclareMathAlphabet{\mathbfit}{OT1}{cmr}{bx}{it}
  \SetMathAlphabet\mathbfit{bold}{OT1}{cmr}{bx}{it}
  \DeclareMathAlphabet{\mathbfss}{OT1}{cmss}{bx}{n}
  \SetMathAlphabet\mathbfss{bold}{OT1}{cmss}{bx}{n}
  \ifAMStwofonts
    \ifCUPmtlplainloaded \else
      \DeclareSymbolFont{UPM}{U}{eur}{m}{n}
      \SetSymbolFont{UPM}{bold}{U}{eur}{b}{n}
      \DeclareSymbolFont{AMSa}{U}{msa}{m}{n}
      \DeclareMathSymbol{\upi}{0}{UPM}{"19}
      \DeclareMathSymbol{\umu}{0}{UPM}{"16}
      \DeclareMathSymbol{\upartial}{0}{UPM}{"40}
      \DeclareMathSymbol{\leqslant}{3}{AMSa}{"36}
      \DeclareMathSymbol{\geqslant}{3}{AMSa}{"3E}
    \fi
  \fi
\fi 

\ifCUPmtlplainloaded \else
  \ifAMStwofonts \else 
    \def\upi{\pi}
    \def\umu{\mu}
    \def\upartial{\partial}
  \fi
\fi

\title[Microlensing in Q2237+0305]{Microlensing induced
spectral variability in Q2237+0305}
\author[G. F. Lewis {\it et al.}]
       {G. F. Lewis$^{1,2}$\thanks{Present Address: 
Dept. of Physics \& Astronomy, University of
Victoria, P.O.Box 3055, Victoria, B.C., V8W~3P6, Canada and
Dept. of Astronomy, University of Washington, Box 351580,
Seattle, WA 98195, U.S.A.  
Email: {\bf \tt gfl@uvastro.phys.uvic.ca}},
M. J. Irwin$^3$, P. C. Hewett$^1$ and C. B. Foltz$^4$ \\
$^1$Institute of Astronomy, Madingley Road, Cambridge, CB3 0HA \\
$^2$Astronomy Group, Earth \& Space Sciences, SUNY at Stony Brook, 
NY 11794, USA \\
$^3$Royal Greenwich Observatory, Madingley Road, Cambridge, CB3 0EZ \\
$^4$MMT Observatory, University of Arizona, Tuscon, AZ 85721, USA
}
\date{Submission Draft}
\pubyear{1996}

\begin{document}

\maketitle

\begin{abstract}
We present both photometry and spectra of the individual images of the
quadruple gravitational lens system Q2237+0305.  Comparison of spectra
obtained at two epochs, separated by $\sim~3\,$years, shows evidence
for significant changes in the emission line to continuum ratio of the
strong ultraviolet CIV~$\lambda$1549, CIII]~$\lambda$1909 and
MgII~$\lambda$2798 lines. The short, $\sim~1\,$day, light--travel time
differences between the sight lines to the four individual quasar
images rule out any explanation based on intrinsic variability of the
source. The spectroscopic differences thus represent direct detection
of microlensing--induced spectroscopic differences in a quasar. The
observations allow constraints to be placed on the relative spatial
scales in the nucleus of the quasar, with the ultra--violet continuum
arising in a region of $\la~0.05~{\rm pc}$ in extent, while the broad
emission line material is distributed on scales much greater than
this.
\end{abstract}


\section{Introduction}

Fluctuations in the brightness of cosmologically distant sources due
to the action of gravitational microlensing is now a well established
observational phenomenon~\cite{irwin1989,corrigan1991}, also occurring,
spectacularly, within our own Galaxy (cf Alcock et al 1996). Utilizing
observations of the quadruple gravitational lens Q2237+0305 we present
evidence of another manifestation of microlensing; modification of the
observed spectral characteristics of a high--redshift quasar due to
differential amplification of source structures with different spatial
scales by objects of stellar mass along the line-of-sight.

Section~\ref{microlensing} presents an overview of microlensing and
the properties of the quadruple lens Q2237+0305 that make it an almost
ideal target for microlensing investigations.  The spectroscopic and
photometric observations of the Q2237+0305 system are presented in
Section~\ref{observations}, with a discussion of the photometry
appearing Section~\ref{obs_light}.  The results of the spectral
extraction are presented in Section~\ref{spectral}, and an
interpretation follows in Section~\ref{discussion}. The basic data
reduction and the relatively complex procedure for extracting the
spectra of the individual quasar components is presented in an
appendix to this paper.

\section{Microlensing and Q2237+0305}\label{microlensing}

Discovered during the CfA redshift survey~\cite{huchra1985} the
Q2237+0305 system consists of four images of a background quasar,
$z\sim1.69$, separated by $\la~2\scnd0$~\cite{yee1988}. The images are
centred on the core of the lensing galaxy, a bright nearby,
$z\sim0.04$, barred spiral.  The galaxy light in the inner regions of
the Q2237+0305 system is dominated by stars in the central bulge of
the galaxy~\cite{yee1988}, and the galaxy surface brightness profile
was modeled using a de Vaucoulers (1953) ${\rm r^{\frac{1}{4}}}$
profile, with a core radius of $r_e = 7\scnd0$, and an ellipticity of
$\epsilon = 0.31$~\cite{racine1991}.  The stellar density along the
sight-lines to the four quasar images is considerable and microlensing
by stars within the galaxy is expected to significantly perturb the
observed image fluxes.

The microlensing scale-length for an isolated star of mass ${\rm M}$,
its Einstein radius $\eta_0$, is defined to be;
\begin{equation}
\eta_o = \sqrt{ \frac{\rm 4 G M }{\rm c^2 }
\frac{ \rm D_{os} D_{ls} }{ \rm D_{ol} } } ,
\label{ein_rad}
\end{equation}
where ${\rm D_{ij}}$ are angular diameter distances between the
observer (o), lens (l) and source (s)~\cite{schneiderbook}.  This
length is defined in the ``source plane'' at a distance ${\rm D_{os}}$
from an observer.  A point-like source passing within this projected
radius of a microlensing mass is amplified by a factor of at least
$1.34$, although the angular scale of multiple imaging induced by such
lenses is $\sim~10^{-6}{''}$, far below present detection limits
with optical telescopes. In practice, therefore, observations of
microlensed systems are confined to the monitoring the image
brightness fluctuations.

The light curve induced by the passage of an isolated microlens
possesses a very simple form, with a characteristic time-scale,
$\tau$:
\begin{equation}
\tau \equiv \frac{ \eta_o }{\rm V_{eff} } ,
\label{timescale}
\end{equation}
where ${\rm V_{eff} }$ is the effective velocity of the source across
the source plane~\cite{kayser1986}. In such simple cases $\tau$ can
be used to characterize the mass of the
microlenses~\cite{wambsganss1992}.

In a high optical depth regime, as is the case for Q2237+0305, the
action of the individual lensing masses combine in a highly non-linear
fashion, and the time-scale of individual ``events'' no longer
reflects the time-scale given by Equation~\ref{timescale}. Instead, the
light curve of a background source exhibits complex variability
behaviour, including asymmetric
fluctuations~\cite{paczynski1986a,wambsganss_thesis,lewis1993}. 

Notwithstanding the complexities arising from the high optical depth
of stars through the lensing galaxy, Q2237+0305 is unique among known
gravitational lens systems in that the lens is located very close to
the observer and thus the light--travel time differences to the four
images are only $\sim1\,$day.  Intrinsic variations in the luminosity
of the quasar thus manifest themselves in all four images within
$\sim1\,$day and differential variability between images with
time-scales $\gg1$day reflects the effects of microlensing. It is this
effective decoupling of the observed time-scales for intrinsic
variability and microlensing by stars in the galaxy that allowed
photometric monitoring of Q2237+0305 to produce the first detection of
a microlensing event~\cite{irwin1989}.

Chang (1984) demonstrated that the amplification of a source is
dependent on its size relative to the Einstein radius of the
characteristic lensing mass. At caustic crossings a point source is
amplified by an infinite amount, but any physical extent leads to a
finite amplification. Sources whose scale size is only a fraction of an
Einstein radius can be amplified by large factors, while sources whose
scale is much greater than an Einstein radius are amplified by a
negligible amount.
 
For Q2237+0305 the Einstein radius (Equation~\ref{ein_rad}) for a star
of Solar mass, projected into the source plane, is
\begin{equation}
\eta_0 \sim 0.05 h_
{\scriptscriptstyle{50}}^{\scriptscriptstyle{-\frac{1}{2}}}
\parsec .
\label{inparsecs}
\end{equation}
Through--out we assume a standard Freidmann-Walker Universe, with an
$\Omega=1$ of matter distributed smoothly. The cosmological constant,
$\Lambda$, is assumed to be zero.  Within the framework of the
``standard model'' for the central regions of quasars~\cite{rees1984}
this microlensing scale is smaller than the extent of the broad
emission line region, $0.1$--$1.0\parsec$, but significantly in excess
of the size of the ultraviolet--optical continuum-producing region,
$\simlt 10^{-3}\parsec$. Thus, differential microlensing amplification
of the quasar continuum and emission line producing regions by objects
of stellar mass within the lensing galaxy of the Q2237+0305 system is
expected.  This should be observable as a time--dependent variation in
the relative strengths of the broad emission lines and the continuum,
and should also be coupled with microlensing--induced photometric
variability. Such observations would provide verification of the
microlensing hypothesis and provide a powerful diagnostic of quasar
structure on extremely small scales.

\section{Observations}\label{observations}

Quasi--simultaneous spectroscopic and imaging observations of
Q2237+0305 were obtained at two epochs, separated by approximately three
years, using the 4.2m William Herschel Telescope (WHT) at the Roque de
los Muchachos Observatory, La Palma.

\subsection{CCD Imaging}

Direct images of Q2237+0305 were obtained at both epochs through a
KPNO broad--band R filter using the auxiliary port at the Cassegrain
focus of the WHT. On 1991 August 12, between 04:00 and 04:40 UT, six
$300\,$s exposures using an EEV CCD were acquired. The seeing was
excellent with a measured FWHM = $0\scnd5$. The second epoch images
consisted of a $300\,$s exposure obtained on 1994 August 14, between
23:50 and 23:54 UT, also employing an EEV CCD. Again, atmospheric
conditions were excellent with the seeing measured to have a FWHM =
$0\scnd4$.  The image scale with the EEV CCD at the auxiliary port of
the WHT is $0\scnd11$/pixel.  Bias frames, and an exposure of a region
of sky devoid of bright stars for flat-fielding were obtained during
evening twilights. Bias subtraction and flat-fielding procedures were
undertaken using {\tt IRAF}~\footnote{{\tt IRAF} is distributed by the
National Optical Astronomy Observatories, which are operated by AURA
under co-operative agreement with the National Science Foundation.}
routines.

\begin{figure*}
\centerline{
\psfig{figure=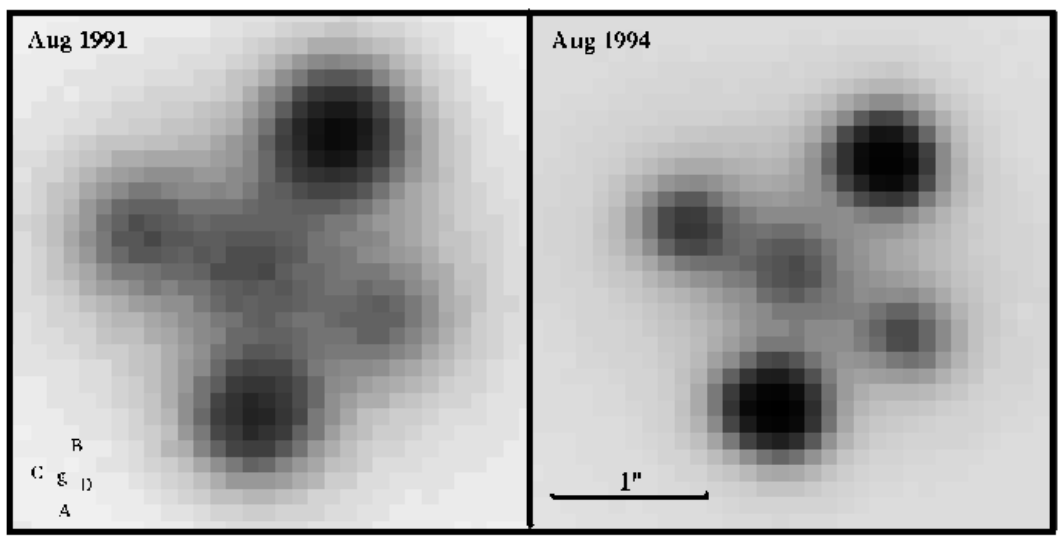,width=5.5in,angle=270}
}
\caption[]{ 
A grey--scale representation of the R--band WHT images of Q2237+0305
from the 1991 and 1994 epochs.  The images have identical scales and
orientations, and show a region $\sim3\scnd25$ on a side with north to
the top and east to the left.  The legend in the lower left--hand
corner of the August 1991 images shows the image designations of Yee
(1988). Image A can be seen to have brightened substantially over the
three year period.}
\label{qso:aux}
\end{figure*}

\subsection{Spectroscopy: Instrumental Configuration}

The spectroscopic observations were acquired with the ISIS
double--beam spectrograph~\cite{isis_manual} at the Cassegrain focus
of the WHT. On 1991 August 14 the spectrograph was configured with the
$5400$ dichroic, splitting the red and blue beams at $\sim
5400$\AA. The R158B grating in the blue arm combined with a thick
$1152\times1242$ pixel EEV CCD with 22.5${\rm \mu m}$ pixels produced
a dispersion of 2.70~\AA$/{\rm pixel}$. The grating angle was set to
give a central wavelength of $\sim 4500$\AA. The wavelength coverage
in the blue is truncated at $\sim 3500$\AA~by the combined effects of
atmospheric absorption and the rapidly falling sensitivity of the CCD,
and in the red, at $\sim 5500$\AA, by the dichroic cross--over
wavelength. The R316R grating in the red arm combined with a second
thick $1152\times1242$ pixel EEV CCD with 22.5${\rm \mu m}$ pixels
produced a dispersion of 1.40~\AA$/{\rm pixel}$. The central
wavelength was set at $\sim 7800$\AA, giving a wavelength coverage
$\lambda\lambda 7000$--$8600$\AA.  The spatial scale on the detector
along the slit was $0\scnd335$/pixel for both red and blue arms.

The 1994 August 17 observations were obtained with a very similar
instrumental configuration. Both red and blue gratings, central
wavelengths and wavelength coverage were as for the 1991 observations.
The $5700$ dichroic split the red and blue beams at $\sim 5700$\AA,
changing the wavelength coverage between the two arms slightly. The
blue--arm detector had been upgraded to a $1024\times1024$ pixel
thinned TEK CCD with 24${\rm \mu m}$ pixels, decreasing the dispersion
to 2.88~\AA$/{\rm pixel}$ but producing much improved sensitivity.  The
spatial scale on the detectors along the slit was $0\scnd335$/pixel in
the red arm and $0\scnd358$ in the blue arm.

Observations at both epochs therefore covered wavelengths
$\lambda\lambda 3500-5400$ and $\lambda\lambda 7000-8600$, which, at
the redshift of $z=1.69$ for Q2237+0305, include the prominent
rest--frame emission lines of CIV $\lambda$1549, CIII] $\lambda$1909
and MgII $\lambda$2798 at wavelengths of $\sim 4175$\AA, $5145$\AA \
and $7540$\AA, respectively.  Unfortunately, the strong telluric
absorption band at $\sim 7600$\AA \ coincides with the red wing of the
MgII emission line.

\subsection{Spectroscopy: Observations}

The atmospheric conditions during the spectroscopic observations at
both epochs (1991 August 15 00:00-05:00UT and 1994 August 17
05:00-05:30UT), were excellent. The seeing derived from the spatial
extent of the component quasar images was measured to be
$\simlt~0\scnd8$ at both epochs. A slit-width of $0\scnd70$ was used
in 1991, producing spectral resolutions (FWHM of an unresolved arc
line) of 8.1\AA~and 4.1\AA~in the blue and the red arms
respectively. In 1994 a slit-width of $0\scnd91$ was employed giving
spectral resolutions of 5.7\AA, ~and 4.1\AA~in the blue and the red
arms respectively.  Q2237+0305 was observed either side of the
meridian with spectra obtained at airmasses of 1.1 to 1.4. The lower
resolution in the blue--arm during the 1991 observations is probably a
result of non--optimal spectrograph focus.

Spectra of pairs of images on the same side of the galaxy nucleus were
obtained in order to minimize contamination from the galaxy core.
Observations of sufficient pairs were acquired to ensure some
redundancy in the observations of each component. The longer 1991
series comprised image pairs A+D, B+C, B+D, and A+C while the short
sequence in 1994 (obtained as a Service allocation) comprised A+C and
B+C. The four individual quasar images were clearly visible on the
acquisition TV and the spectrograph slit, oriented to the appropriate
position angle for the particular image pair, was aligned across each
image pair using visual inspection. A log of the spectroscopic
observations is given in Table~\ref{table_obs}. It was necessarily
impossible to acquire the spectra of the image pairs with the
spectrograph slit aligned along the parallactic angle.  Differential
atmospheric dispersion was thus expected to produce a systematic loss
of light from the slit at the bluest wavelengths.

\begin{table}
\begin{center}
\begin{tabular}{|c|c|c|c|c|c|}
\hline
Year & Image Pair & Slit PA & Exposure & Airmass & Seeing\\ \hline
1991 & A+D & 302.8  & 1800 & 1.386 & $0\scnd80$ \\
1991 & A+D & 302.8  & 1800 & 1.254 & $0\scnd80$ \\
1991 & B+C & 290.8  & 1800 & 1.144 & $0\scnd80$ \\
1991 & B+C & 290.8  & 1800 & 1.125 & $0\scnd80$ \\
1991 & B+D &   8.8  & 1800 & 1.111 & $0\scnd80$ \\
1991 & B+D &   8.8  & 1800 & 1.134 & $0\scnd80$ \\
1991 & A+C &  27.8  & 1800 & 1.218 & $0\scnd80$ \\
1991 & A+C &  27.8  & 1800 & 1.316 & $0\scnd80$ \\ 
1994 & A+C &  27.8  & 1500 & 1.109 & $0\scnd65$ \\
1994 & B+C & 290.8  & 1500 & 1.108 & $0\scnd65$ \\ \hline
\end{tabular}
\caption[]{
Journal of spectroscopic observations of Q2237+0305 taken on the
nights of 1991 August 14/15 and 1994 August 17/18.} 
\label{table_obs}
\end{center}
\end{table}

Bias frames and exposures of quartz and tungsten lamps were obtained at
the start of each night and flux standards~\cite{massey1988,oke1990}
were observed at the end of the night.  Exposures of Cu+Ar (blue arm)
and Cu+Ar/Cu+Ne (red arm) lamps were taken following each exposure of
an image pair or a standard star.

\section{Photometry}\label{obs_light}

Figure~\ref{qso:aux} shows a grey--scale representation of the reduced
R--band frames from the 1991 and 1994 epochs. Relative photometry of
the individual quasar images was performed using the procedures
described in Corrigan~(1993). Exposures of standard stars were not
obtained at either epoch and the frames were calibrated using stars
visible both in frames from Corrigan~(1993) and the new images. The
uncertainty in the calibration is estimated to be
$\pm0.05\,$mag. R--band magnitudes for the four quasar images on the
R--band system used in Corrigan et al (1991), are given in
Table~\ref{table_mags}. The significant difference in the relative
brightness of images A and B between the two epochs is evident from
inspection of Figure~\ref{qso:aux}, and the change in relative
brightness, $\Delta m_{A-B} = 0.38\pm 0.07\,$mag is well determined,
independent of the calibration procedure.

Relative, and calibrated, magnitudes for the images at both epochs are
consistent with the photometric time series (27 epochs) for the period
1990 to 1993, obtained by {\O}stensen (1994) at the Nordic Optical
Telescope, which shows image A brightening to become comparable to
image B at the end of 1992.

\begin{table}
\begin{center}
\begin{tabular}{|c|c|c|c|c|c|}
\hline
       & Seeing    & A    & B    & C    & D    \\ \hline
 1991  & $0\scnd5$ & 17.53 & 17.23 & 18.11 & 18.28 \\
 1994  & $0\scnd4$ & 17.12 & 17.20 & 18.00 & 18.23 \\ \hline
\end{tabular}
\caption[]{
The R-band image magnitudes of Q2237+0305 at the 1991 and 1994
epochs. The errors in these measurements is 0.05 mag.}
\label{table_mags}
\end{center}
\end{table}

\section{Spectroscopic Reductions}\label{spectral}

The relatively complex nature of the data, consisting of two merging
quasar spectra superimposed upon an extended galaxy background,
precluded the application of standard data reduction software to
extract the spectra of the individual quasar components.  A new
reduction technique, therefore, was developed for application to this
data.  This technique is presented, in some detail, in
Appendix~\ref{spectralappendix}.

\subsection{Spectral Fitting}

The image separations, seeing and $\left< \chi^2 \right>$, the average
``goodness-of-fit'' value (Appendix~\ref{gof}), from the
model--fitting procedure applied to the blue and red spectroscopic
observations of each image pair are presented in
Table~\ref{fitting_results}. The seeing is defined as the FWHM of the
Gaussian core of the profile model fitted to the two quasar spectra
(Section~\ref{psf}). For comparison, image separations derived from
Hubble Space Telescope imaging of the system~\cite{rix1992} are
presented in Table~\ref{fitting_results}.

We note that the red spectra in Table~\ref{fitting_results} possess
systematically greater estimates of the width of the seeing profile
when compared to the equivalent blue spectra. This may be due to an
under--estimate of the galaxy component in the wings of the quasar
profile, driving profile width to larger values. This, however, was
not readily apparent in ``by-eye'' inspections of the fits as a
function of wavelength. Such model dependencies will not effect the
over--all conclusions presented in this paper.

\begin{table*}
\begin{center}
\begin{tabular}{|c|c||c|c|c||c|c|c|c|}
\hline
 & & \multicolumn{3}{|c|}{\underline{\makebox[4.5cm]{Blue Spectra}}}& 
\multicolumn{3}{|c|}{\underline{\makebox[4.5cm]{Red Spectra}}}   \\
Images & Year & $\Delta$ & FWHM & $\left< \chi^2 \right>$  & $\Delta$ 
& FWHM & $\left< \chi^2 \right>$ & $\Delta_{\rm HST}$ \\ \hline
A+D & 1991 & $1\scnd07\pm0\scnd25$ & $1\scnd06\pm0\scnd11$ & 1.23 & $1\scnd07\pm0\scnd23$ & $1\scnd13\pm0\scnd14$ & 1.91 & $1\scnd01$ \\  
A+D & 1991 & $1\scnd09\pm0\scnd26$ & $0\scnd94\pm0\scnd11$ & 1.29 & $1\scnd05\pm0\scnd24$ & $1\scnd11\pm0\scnd13$ & 1.91 & $1\scnd01$ \\  
B+C & 1991 & $1\scnd38\pm0\scnd22$ & $0\scnd83\pm0\scnd13$ & 1.34 & $1\scnd44\pm0\scnd19$ & $1\scnd01\pm0\scnd11$ & 1.48 & $1\scnd39$ \\  
B+C & 1991 & $1\scnd36\pm0\scnd25$ & $0\scnd90\pm0\scnd18$ & 1.40 & $1\scnd42\pm0\scnd22$ & $1\scnd05\pm0\scnd13$ & 1.74 & $1\scnd39$ \\  
B+D & 1991 & $1\scnd23\pm0\scnd27$ & $0\scnd72\pm0\scnd05$ & 1.21 & $1\scnd21\pm0\scnd30$ & $1\scnd05\pm0\scnd10$ & 1.77 & $1\scnd18$ \\  
B+D & 1991 & $1\scnd20\pm0\scnd31$ & $0\scnd72\pm0\scnd06$ & 1.25 &     -      &     -      &  -   \\  
A+C & 1991 & $1\scnd33\pm0\scnd22$ & $0\scnd84\pm0\scnd14$ & 1.40 & $1\scnd34\pm0\scnd20$ & $1\scnd12\pm0\scnd13$ & 1.60 & $1\scnd35$ \\  
A+C & 1991 & $1\scnd28\pm0\scnd22$ & $0\scnd89\pm0\scnd11$ & 1.30 & $1\scnd35\pm0\scnd21$ & $1\scnd17\pm0\scnd13$ & 1.61 & $1\scnd35$ \\ 
A+C & 1994 & $1\scnd33\pm0\scnd09$ & $0\scnd57\pm0\scnd03$ & 1.46 & $1\scnd24\pm0\scnd12$ & $0\scnd99\pm0\scnd05$ & 1.60 & $1\scnd35$ \\  
B+C & 1994 & $1\scnd33\pm0\scnd30$ & $0\scnd79\pm0\scnd10$ & 1.06 & $1\scnd34\pm0\scnd13$ & $0\scnd90\pm0\scnd08$ & 1.47 & $1\scnd39$\\  
\hline
\end{tabular}
\caption[Spectral Fitting Parameter]{
\label{fitting_results}
Spatial quantities (image separation and seeing) and average
goodness--of--fit derived from the model--fitting to each of the blue
and red spectroscopic observations. The one--sigma errors are derived
from the total spectral fit.  The final column presents the image
separations, $\Delta_{\rm HST}$, from the HST observations of
Rix~\etal~(1992). The quoted error in the image positions as derived
from HST imaging is $\la0\scnd015$.}
\end{center}
\end{table*}

\subsection{The Spectra}

\begin{figure*}
\centerline{
\psfig{figure=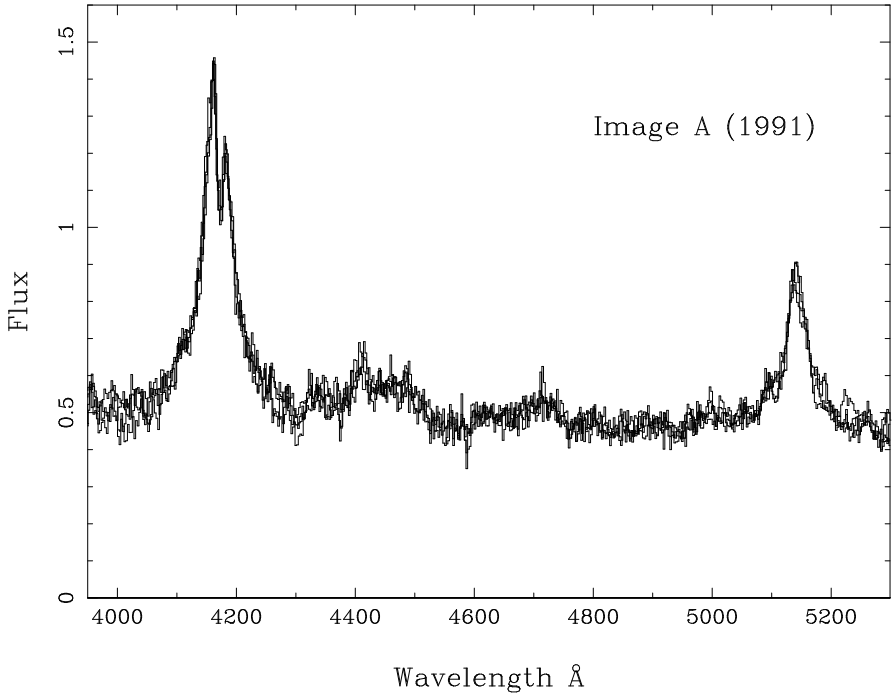,width=6in,angle=270}
}
\caption[Spectrum of Image A]
{The four blue spectra of Image A obtained in 1991. The spectra have
been corrected for the instrument response and normalized such that
the mean flux in the continuum region 4400~$\rightarrow$~4900\AA~is
identical.}
\label{comp_spectra}
\end{figure*}

As an illustrative example of the extracted spectra,
Figure~\ref{comp_spectra} presents all four blue spectra of image A
obtained in 1991. The spectra are normalized such that the average
flux in the continuum region 4400~$\rightarrow$~4900\AA~is the same.
To within the noise, there is excellent agreement between the
independent observations of the spectrum, with typical differences 
being $\sim10\%$.

It should be noted that the spectra show no apparent distortion due to
the effects of atmospheric dispersion. The A+D and A+C image pairs
were observed at very similar zenith distances, but either side of the
upper culmination. At these air masses the resulting slit orientations
lay within $\sim15^o$ of the parallactic angle, resulting similar
atmospheric dispersion for each spectrum. This was not the case for
the 1994 observations, both of which were observed very near the
upper culmination. Atmospheric dispersion, coupled with a narrow slit
and good seeing, produced significant loss of light at the blue end of
the spectra.

\subsection{Scaled Spectra}\label{scaled_spectra}

The narrow slits and the requirement to orient the position
angle away from the parallactic, (Section~\ref{observations}), mean
that spectra of the individual images cannot be compared using an
absolute flux scale.  However, the principal effect predicted to arise
from differential microlensing of the continuum and emission line
producing regions will manifest itself as differences in emission line
equivalent--width among the spectra of the images.

\begin{figure*}
\centerline{
\psfig{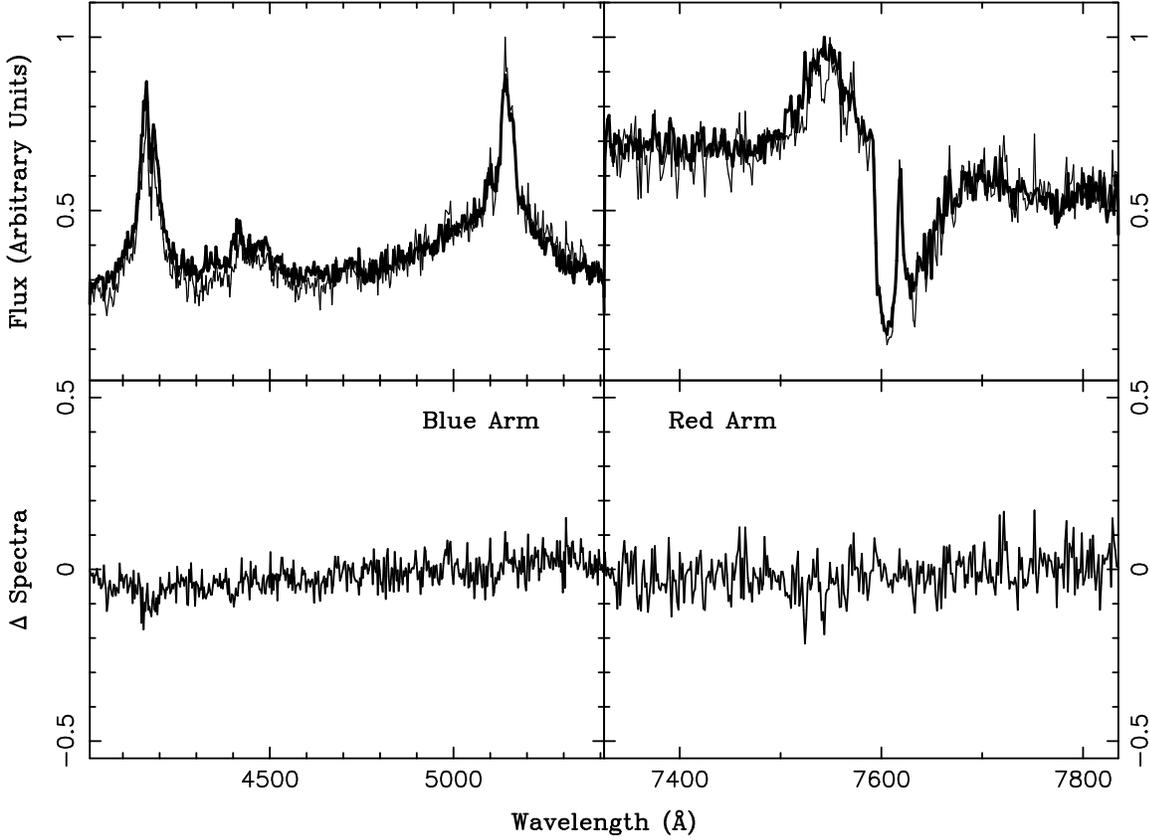}
}
\caption[Scaled Spectra of Images ${\rm A+C}$]
{The blue (left panel) and red (right panel) ISIS spectra for images
${\rm A+C}$ from the 1991 observations.  In the top panels the
spectrum of C has been scaled to match the spectrum of image A in the
$\rm \left( 4650\rightarrow4900 \AA \right)$ in the blue spectrum, and
$\rm \left( 7000\rightarrow7080 \AA \right)$ in the red spectrum.  The
brighter A image is presented as a thick line, where as the scaled C
spectrum is a thin line. The lower panel presents the difference in
these spectra, defined as the spectra of A minus the scaled spectrum
of C.}
\label{scaled_1}
\end{figure*}

\begin{figure*}
\centerline{
\psfig{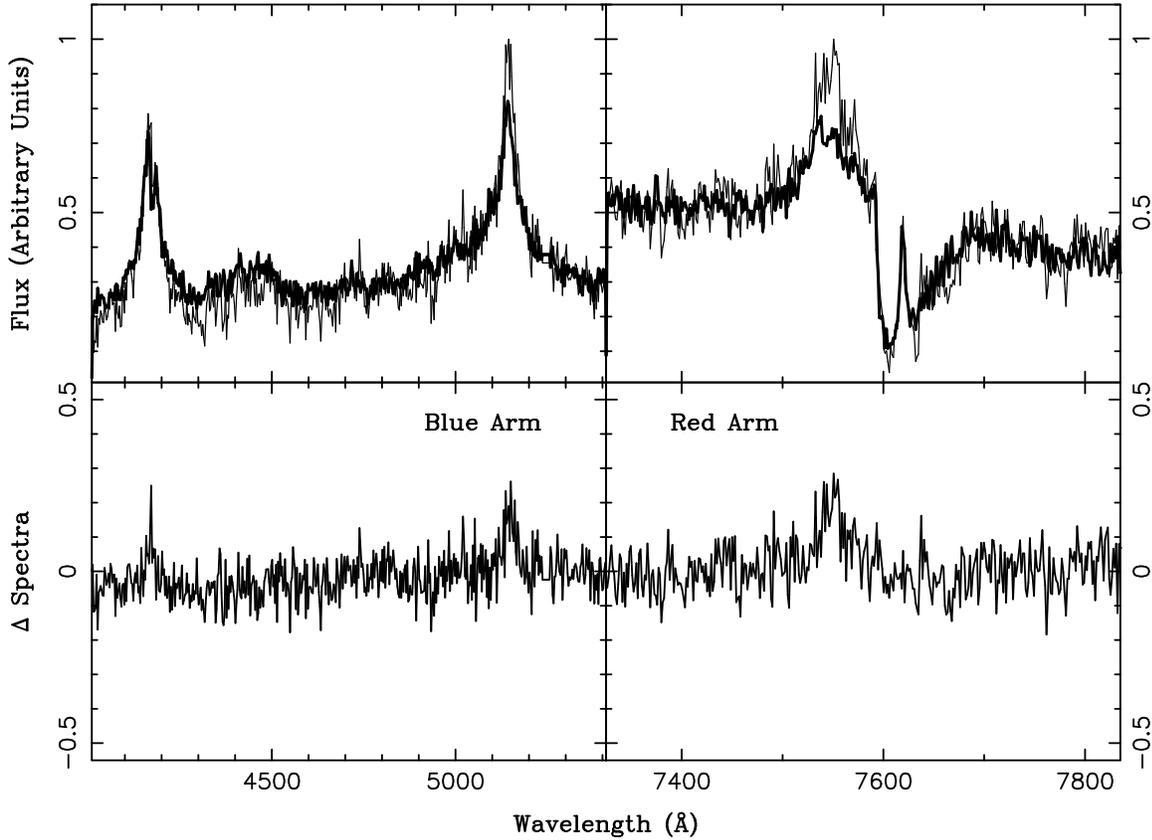}
}
\caption[Scaled Spectra of Images ${\rm A+D}$]
{ As for Figure~\ref{scaled_1}, presenting the spectra for image A and
  image D in 1991, which has been scaled so that the continua match.
  The thinner line represents the scaled spectrum of the D image,
  while the thicker line is for image A. The difference between these
  spectra, presented in the lower panels, is defined to be the
  spectrum of the A image subtracted from the scaled D spectrum.}
\label{scaled_2}
\end{figure*}

\begin{figure*}
\centerline{
\psfig{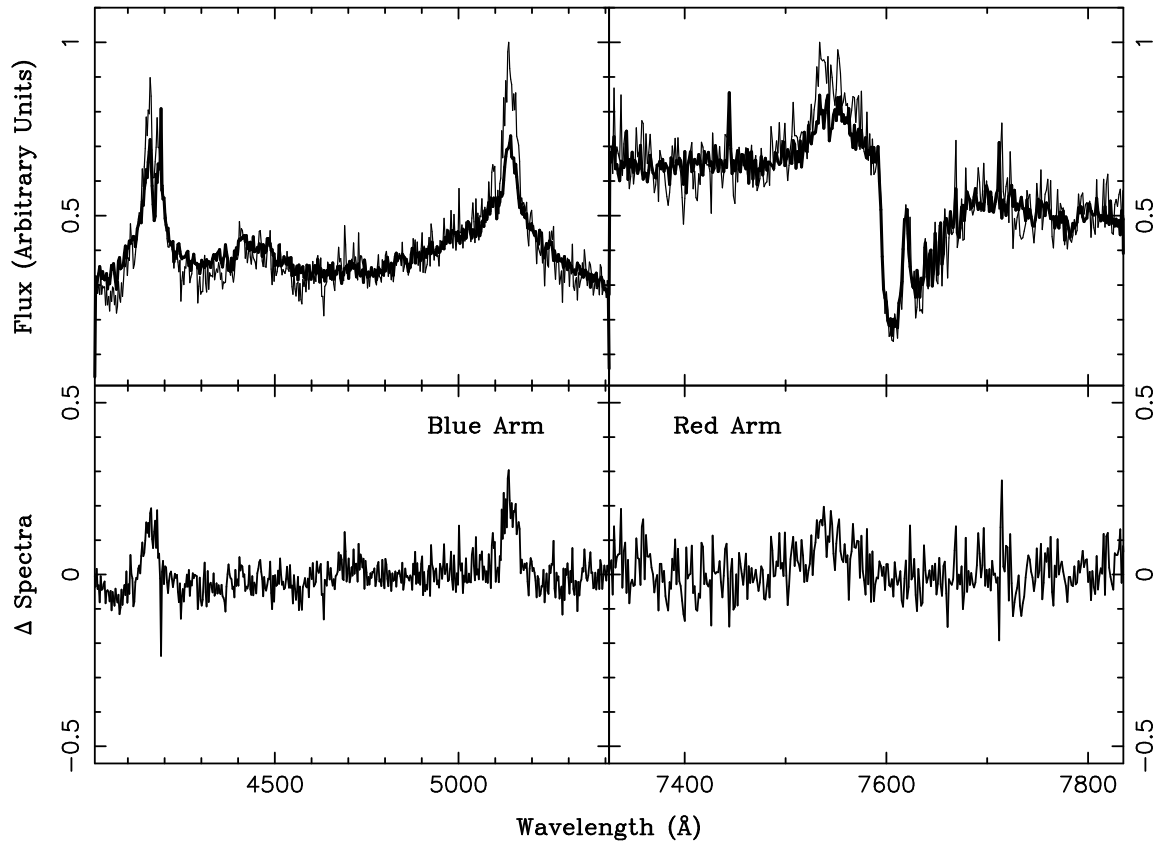}
}
\caption[Scaled Spectra of Images ${\rm B+C}$]
{ As for Figure~\ref{scaled_1}, presenting the spectra for image B and
  image C in 1991, which has been scaled so that the continua match.
  In the top panels, the B image is represented by a thick line,
  whereas the scaled C image is a thin line. The difference between
  these spectra is presented in the lower panel and the spectrum of
  the B image subtracted from the scaled spectrum of the C image.}
\label{scaled_3}
\end{figure*}

\begin{figure*}
\centerline{
\psfig{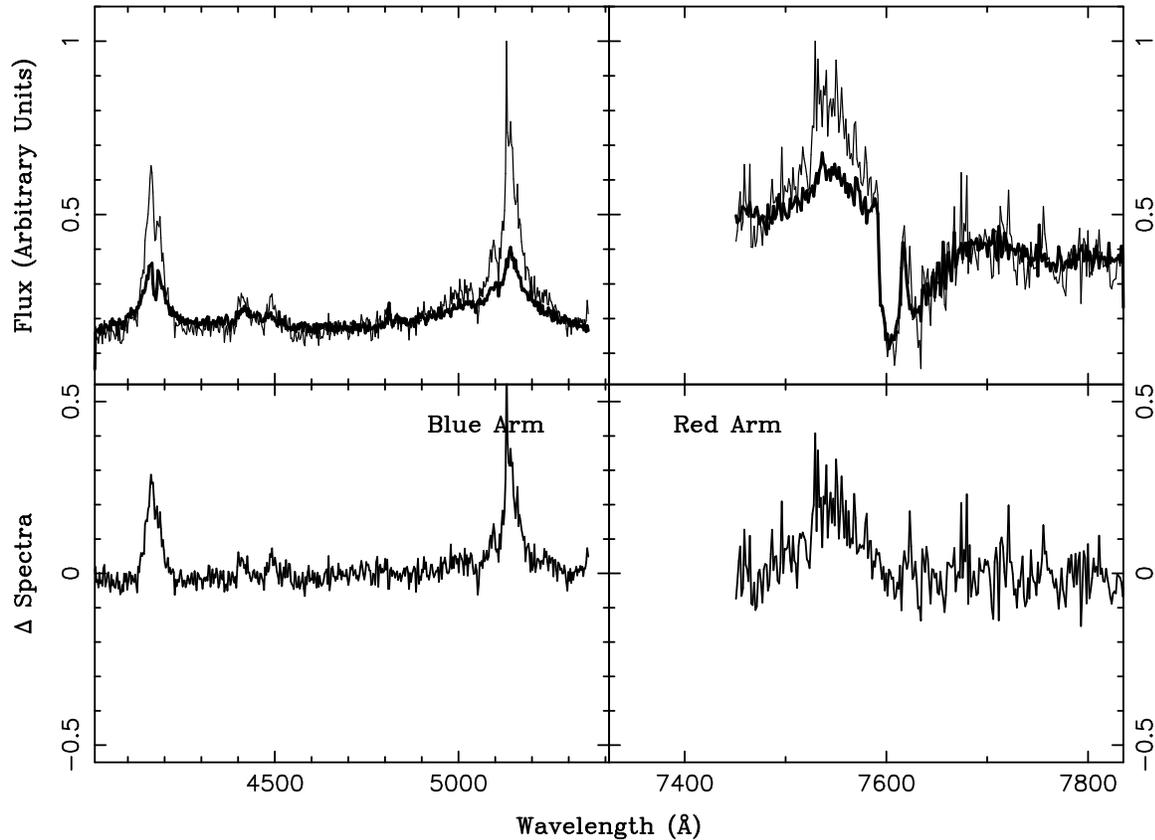}
}
\caption[Scaled Spectra of Images ${\rm B+D}$]
{ As for Figure~\ref{scaled_1}, presenting the spectra for image B and
  image D in 1991, which has been scaled so that the continua match.
  In the upper panels, the B image is represented by a thick line,
  while the scaled D image is a thin line. Although the continua agree
  the relative flux in the emission lines is very different between
  the images. This is also apparent in the lower plots which presents
  the scaled D image minus the B image. The data red--ward of ${\rm
  5350\AA}$ in the left--hand plot, and blue--ward of ${\rm 7400\AA}$
  in the right--hand plot were neglected due to very poor signal in
  the D spectrum at these wavelengths.}
\label{scaled_4}
\end{figure*}

Any such differences in equivalent--widths between the spectra should
be evident visually if the spectra are scaled such that their
continuum levels are coincident. Equivalent--width differences will
then manifest themselves as emission line strength changes.
Figures~\ref{scaled_1}--\ref{scaled_4} show the 1991 epoch blue and
red spectra of such continuum--scaled spectra for the image pairs,
A+C, A+D, B+C and B+D respectively. The spectra are plotted following
the multiplicative scaling of the fainter spectrum such that the mean
fluxes in the ``continuum'' regions 4650~$\rightarrow$~4900\AA~(in the
blue), and 7000~$\rightarrow$~7080\AA~in the red spectra are equal.

Spectra of the image pair A+B (Figure~\ref{scaled_1}) show no obvious
systematic difference in emission line strength. However,
Figure~\ref{scaled_2} shows image D has significantly stronger
emission lines than image A and similarly Figure~\ref{scaled_3}
demonstrates that image C has stronger emission lines than image
B. Consistency in the emission line strength differences between the
images is demonstrated in Figure~\ref{scaled_4} where image D is seen
to possess a very marked excess of emission compared to the spectrum
of image B. In addition to the three prominent emission lines of
C{\footnotesize IV}, C{\footnotesize III]} and Mg{\footnotesize II},
evidence for an excess at 4455\AA, coincident with He{\footnotesize
II} $\lambda1640$ is present.

\subsection{Emission Line Equivalent Widths}

The emission line equivalent width ($EW$) provides a quantitative measure of emission line strength:
\begin{equation}
EW = \int_{\lambda_1}^{\lambda_2} \frac{ \left[ S\left(\lambda\right) 
- I\left(\lambda\right) \right] }{ I\left(\lambda\right) }\ \  d\lambda
\label{ew}
\end{equation}
where $S\left(\lambda\right)$ is the flux at
wavelength $\lambda$and
$I\left(\lambda\right)$ is the flux in the underlying continuum at a
wavelength $\lambda$. Wavelengths, $\lambda_1$ and $\lambda_2$ are
chosen to bound the line of interest.

Establishing the true continuum level in rest--frame ultraviolet
spectra of quasars is difficult due to the presence of weak emission
features [cf the composite quasar spectrum of Francis~\etal~(1991)].
However, establishing the absolute equivalent width is not required,
rather, a consistent measure of equivalent width that allows the
relative emission line strength in the component spectra is all that is
necessary. A continuum was estimated for each emission line based on
the median flux within ``continuum'' bands on either side of the line.
The continuum was set by the straight line connecting points defined by
the median flux in each band at the mid--point of the wavelength range
of each band. The EW was then calculated numerically according to
Equation~\ref{ew} with the summation extending from the long
wavelength edge of the blue continuum band to the short wavelength edge
of the red continuum band (Table~\ref{wavelength_range}). In the case
of the Mg{\footnotesize II} line the summation was truncated at 7590\AA
\ to avoid the strong telluric absorption feature at $\sim 7600$\AA.
 
\begin{table}
\begin{center}
\begin{tabular} {|c|c|c|c|c|}
\hline
   Line & $x_{l,1}$ & $x_{l,2}$ & $x_{u,1}$ & $x_{u,2}$ \\ 
\hline 
C{\footnotesize IV}  & 3970  & 4040 & 4290 & 4300   \\
C{\footnotesize III]}& 5020  & 5060 & 5220 & 5260  \\
Mg{\footnotesize II} & 7450  & 7480 & 7700 & 7739  \\
\hline
\end{tabular}
\caption[Wavelength Range for the Equivalent Width Measurement]
{\label{wavelength_range} For the emission lines, $x_l$ and
  $x_u$ represent the bounds on the regions used in the estimation of
  the continuum. The equivalent width is measured between $x_{l,2}$ and
  $x_{u,1}$, except for Mg{\footnotesize II} where the summation was
  truncated at 7590\AA~to exclude the atmospheric A--band absorption.}
\end{center}
\end{table}

\begin{table}
\begin{center}
\begin{tabular}{|c|c||c|c|c|}
\hline
Image & Year & C{\footnotesize IV} & C{\footnotesize III]} & Mg{\footnotesize II} \\ \hline
A & 1991 & $126.11\pm1.52$ & $43.73\pm1.34$ & $28.65\pm0.96$ \\
D & 1991 & $190.53\pm4.35$ & $61.28\pm2.96$ & $49.83\pm1.92$ \\ 
A & 1991 & $132.23\pm1.72$ & $38.78\pm1.12$ & $33.10\pm0.92$ \\
D & 1991 & $203.47\pm5.08$ & $64.02\pm2.64$ & $43.09\pm1.72$ \\ 
B & 1991 & $68.32\pm0.65$  & $20.91\pm0.89$ & $18.08\pm0.53$ \\
C & 1991 & $116.87\pm2.06$ & $47.23\pm2.41$ & $28.53\pm1.18$ \\ 
B & 1991 & $77.00\pm0.61$  & $25.66\pm0.86$ & $19.65\pm0.55$ \\
C & 1991 & $120.34\pm1.93$ & $45.66\pm2.30$ & $32.61\pm1.17$ \\ 
B & 1991 & $69.59\pm0.68$  & $28.96\pm0.93$ & $19.26\pm0.64$ \\
D & 1991 & $148.88\pm2.65$ & $83.12\pm2.11$ & $49.23\pm1.92$ \\ 
B & 1991 & $69.70\pm0.84$  & $29.03\pm0.91$ &       -        \\
D & 1991 & $179.62\pm3.73$ & $93.29\pm3.31$ &       -        \\ 
A & 1991 & $99.91\pm1.03$  & $40.61\pm1.06$ & $28.05\pm0.73$ \\
C & 1991 & $112.41\pm2.12$ & $36.92\pm1.76$ & $25.05\pm1.16$ \\ 
A & 1991 & $118.33\pm1.33$ & $38.20\pm1.16$ & $27.48\pm0.77$ \\
C & 1991 & $83.37\pm2.20$  & $37.45\pm2.01$ & $24.76\pm1.29$ \\ 
A & 1994 & $102.21\pm0.43$ & $36.44\pm0.63$ & $25.93\pm0.46$ \\
C & 1994 & $151.80\pm1.60$ & $51.73\pm1.64$ & $29.96\pm0.94$ \\ 
B & 1994 & $101.45\pm1.12$ & $30.51\pm1.64$ & $17.36\pm0.58$ \\
C & 1994 & $141.34\pm3.18$ & $45.51\pm1.93$ & $20.61\pm1.10$ \\ \hline
\end{tabular}
\caption[Equivalent Width Measurements]{
\label{ew_table}
Equivalent width measurements for images at both epochs. The error estimate
is calculated assuming a continuum placement uncertainty of 2\% for
C{\footnotesize IV} and C{\footnotesize III}, and a 5\% error for
Mg{\footnotesize II}.  }
\end{center}
\end{table}

\begin{table}
\begin{center}
\begin{tabular}{|c|c||c|c|c|}
\hline
Image & Year & C{\footnotesize IV} & C{\footnotesize III]} & Mg{\footnotesize II} \\ \hline
A & 1991 &  $119.14\pm12.15$ & $41.33\pm3.80$  & $29.32\pm2.22$ \\
B & 1991 &  $71.15\pm3.42$   & $26.14\pm3.31$  & $19.00\pm0.67$ \\
C & 1991 &  $108.33\pm14.50$ & $41.82\pm4.67$  & $27.74\pm3.18$ \\
D & 1991 &  $180.63\pm20.18$ & $75.43\pm13.31$ & $47.38\pm3.08$ \\  
A & 1994 &  $102.21$         & $36.44$         & $25.93$        \\
B & 1994 &  $101.45$         & $30.51$         & $17.36$        \\
C & 1994 &  $146.57\pm5.23$  & $48.62\pm3.11$ & $25.28\pm4.67$ \\ \hline
\end{tabular}
\caption[Mean Equivalent Width Measures]{
\label{mean_ew}
Mean values of the emission line equivalent widths from independent
spectra. The $1\sigma$ errors represent the measured error in the
mean.  Only one spectrum of each of images A and B was obtained in
1994, and the errors in the values of the equivalent widths are
assumed to be similar for those found for Image C.}
\end{center}
\end{table}

Table~\ref{ew_table} presents equivalent width measurements of
C{\footnotesize VI}, C{\footnotesize III]} and Mg{\footnotesize II} for
spectra taken in 1991 and 1994. The errors are calculated assuming an
uncertainty in the continuum placement of 2\% for C{\footnotesize IV}
and C{\footnotesize III}, and 5\% for Mg{\footnotesize II}. Mean
equivalent widths for each image at each epoch are given in
Table~\ref{mean_ew}, together with the $1\sigma$ errors in the mean.

\subsection{Emission Line Centroid Shifts}

Nemiroff~(1988) and Schneider and Wambsganss~(1990) predicted that
gravitational microlensing of substructure in the broad line emitting
region would result in significant profile differences between the same
emission line in different images. Selective microlensing enhancement
of portions of the emission line region undergoing ordered rotation, or
outflow, could result in shifts of $\sim1000$\kms~in the central
wavelengths of emission lines.

Any differences in emission line centroids can be quantified by
cross--correlation. The cross--correlation between two discrete signals
$\mu_i$ and $\nu_j$ is defined to be:
\begin{equation}
C(\mu,\nu)_k \equiv \sum_{j=0}^{N-1} \mu_{k+j} \nu_j .
\label{cross_cor}
\end{equation} 
Computation of the cross--correlation function using observed galaxy
spectra and zero--redshift templates is a standard technique in
establishing redshifts~\cite{tonry1979}. Here, emission lines in
different images arising from the same transition were
cross--correlated. The quasar continuum was first subtracted from each
emission line using the continuum definition procedure employed to
perform the measurement of the equivalent widths. Then, the pairs of
continuum subtracted emission lines were cross--correlated using the
Numerical Recipes {\tt CORREL} routine~\cite{num_rec}.  Due to the
presence of the telluric A--band absorption, only the region blue--ward
of 7590\AA~was used in the analysis of the Mg{\footnotesize II}
velocity characteristics.

A Gaussian function was fitted to the central portion of the resulting
cross--correlation function in order to provide a measure of the peak
location. The function took the form:
\begin{equation}
C(x) = A e^{ - \left( \frac{ \left| x - x_c \right| }{ c } 
\right)^{ 2 d }}
\label{super_gauss}
\end{equation}
where $x_c$ is the central position of the peak of the distribution of
width $c$. $A$ is a normalization factor, and $d$ is a free parameter.
The fitting algorithm was an implementation of the Levenberg-Marquardt
minimization method, as presented by Press~\etal~(1988) in {\tt
MRQMIN}. The peak--fitting was undertaken using apertures ranging from
ten to eighty pixels either side of the peak. The location of the peak
centre depended somewhat upon the aperture used, as the form of the
cross--correlation peak is not perfectly represented by
Equation~\ref{super_gauss} ie large apertures tended to fit the
cross--correlation peak wings, while small apertures didn't possess
the curvature sufficient to accurately identify the peak.  For these
reasons various apertures were applied to each cross-correlation
peak. This resulted in the selection of a separate aperture to to
reproduce the form of each peak.

The sensitivity of the cross--correlation analysis with spectra of the
signal--to--noise ratio available is approximately $50\kms$
($1\sigma$).  The velocity shifts for any of the three emission lines
between any of the image pairs were all consistent with no velocity
offset.  For each line in the 1991 data, the mean velocity was;
$4.6\pm20.9\kms$ for C{\footnotesize IV}, $48.1\pm88.9\kms$ for
C{\footnotesize III} and $18.0\pm27.3\kms$ for Mg{\footnotesize
II}. The uncertainties quoted are $1\sigma$ and are based on the
uncertainty of the peak--fitting procedure described above.  No
systematic offset was seen for any of the cross--correlation peaks and
these values suggest that there are no significant differences in the
central velocity of the lines.

Consideration was also given to whether emission line profile
differences were present between images. Such differences are
expected due to some preferential amplification of any substructure 
in the broad emission line region~\cite{nemiroff1988,schneider1990}.
However, the limited signal-to-noise ratio of the data
precludes the derivation of astrophysically useful constraints on
possible line shape differences.

\section{Discussion}\label{discussion}

\subsection{Quasar Emission Regions}

Table~\ref{mean_ew} and the spectra shown in
Section~\ref{scaled_spectra} demonstrate that during the 1991
observations, strong equivalent width differences existed between the
spectra of the four images. This result is consistent with standard
models of quasar structure, with the scale of the broad line emitting
region significantly greater than that of the continuum source.
Schemes, such as the star-burst model, where emission lines originate in
narrow supernovae shells embedded in a large continuum emitting
source~\cite{cid_thesis}, cannot account for the systematic lack of
enhancement of the line flux during microlensing.

\begin{table*}
\begin{center}
\begin{tabular}{|c||c|c|c|c||c|c|c|c|}
\hline
Image & Year & C{\footnotesize IV} & C{\footnotesize III]} & Mg{\footnotesize II} & Year & C{\footnotesize IV} & C{\footnotesize III]} & Mg{\footnotesize II} \\
\hline
A & 1991 & 1.10 & 0.99 & 1.06 & 1994 & 0.70 & 0.75 & 1.02 \\
B & 1991 & 0.66 & 0.63 & 0.68 & 1994 & 0.69 & 0.63 & 0.69 \\
C & 1991 & 1.00 & 1.00 & 1.00 & 1994 & 1.00 & 1.00 & 1.00 \\
D & 1991 & 1.67 & 1.80 & 1.71 &   -  &   -  &  -   &  -   \\ \hline
\end{tabular}
\caption[Normalized Equivalent Widths]
{\label{norm_ew} The equivalent widths for the images in Q2237+0305,
normalized with respect to measurements for image C at each epoch. 
}
\end{center}
\end{table*} 

Table~\ref{norm_ew} presents the equivalent widths from both epochs of
observation, normalized with respect to the equivalent of each line in
the spectrum of image C. During the 1991 observations image A had
equivalent widths of order unity when compared to those of image C,
while each value in image B was $\sim0.65$ that of C. Image D, on the
other had equivalent width measures of $\sim1.7$ those of C, reflecting
the microlensing deamplification of its continuum.  This interpretation
is consistent with recent observations which show image D to have a
radio brightness which is comparable to the other images in this
system~\cite{falco1996}.  The photometry of system at the two epochs
illustrate that image A had brightened over the three years, becoming
as bright as image B. This is reflected in the values of
Table~\ref{norm_ew}. Essentially, the equivalent widths of image B,
relative to those of image C, are unchanged between 1991 and 1994,
while image A, on the other hand, has shown a strong decrease in the
line-to-continuum strength in both C{\footnotesize IV} and
C{\footnotesize III]}. The EW measurements can be explained if image A
underwent a microlensing event between the two epochs. The measure of
the Mg{\footnotesize II} equivalent width in image A, relative to that
of image C, does not change between the 1991 and 1994 observation.  The
reason for this constant value in Mg{\footnotesize II} is still
unknown.

The ratio of the equivalent widths presented in Table~\ref{norm_ew}
indicate that, at the time of observation, the spectral slope of the
quasar images was independent of the level of microlensing
amplification, with an equal level of enhancement in the red and blue
regions of the spectrum. In the rest frame of the quasar, the spectra
presented here cover the range 1500\AA~to 2800\AA, which, if one
adopts a temperature structure of the form $T(R) \propto
R^{-\frac{3}{8}}$ as appropriate for the inner, radiation--dominated
region of an optically thick, geometrically thin accretion
disk~\cite{shakura1973}, implies that
\begin{equation}
\frac{ R_{{\rm 2800}}} {R_{\rm 1500}} \sim 5.
\end{equation}
For the corresponding effective temperatures, these emission regions
are $\la 10^{15}$cm in extent, both the red and blue light comes from
a fraction of an Einstein radius (Equation~\ref{ein_rad}), and
light curves in both colours should exhibit similar properties. The
difference in scale size will only be apparent as colour differences
when a caustic passes over the source~\cite{wambsganss1991} and it is
probable that all the sources lie in inter-caustic regions.

\subsection{Structure in the Broad Line Region}

To the limit of the sensitivity, $\sim 150\kms$, no significant velocity
differences between the same emission lines in the spectra of different
images in the Q2237+0305 system were found. Thus, any substructure in
the broad emission line region must be large enough to be insensitive to
the effects of microlensing. With Equation~\ref{inparsecs}, this
implies substructure in the broad line region must be on scales $\ga
0.05 h_{\scriptscriptstyle{50}}^{\scriptscriptstyle{-\frac{1}{2}}}$
\parsec.

\vfil

\section*{Acknowledgments}

G.F.L. would like to thank Roger Blandford and Bob Carswell for
interesting discussions on the topics of microlensing and quasar
structure. G.F.L. would also like to thank Rachel Webster and the
Astronomy Group at the University of Melbourne for their hospitality,
and Noriaki Yahata for help with some graphics. C.B.F. acknowledges
the support of the NSF through grant AST 93-20715.  A NATO
Collaborative Research Grant (held by P.C.H.) aids research on
gravitational lenses and quasar surveys at the Institute of Astronomy.
The anonymous referee is thanked for constructive comments.

\vfil

\appendix

\section{Spectroscopic Reductions}\label{spectralappendix}

\begin{figure}
\centerline{
\psfig{figure=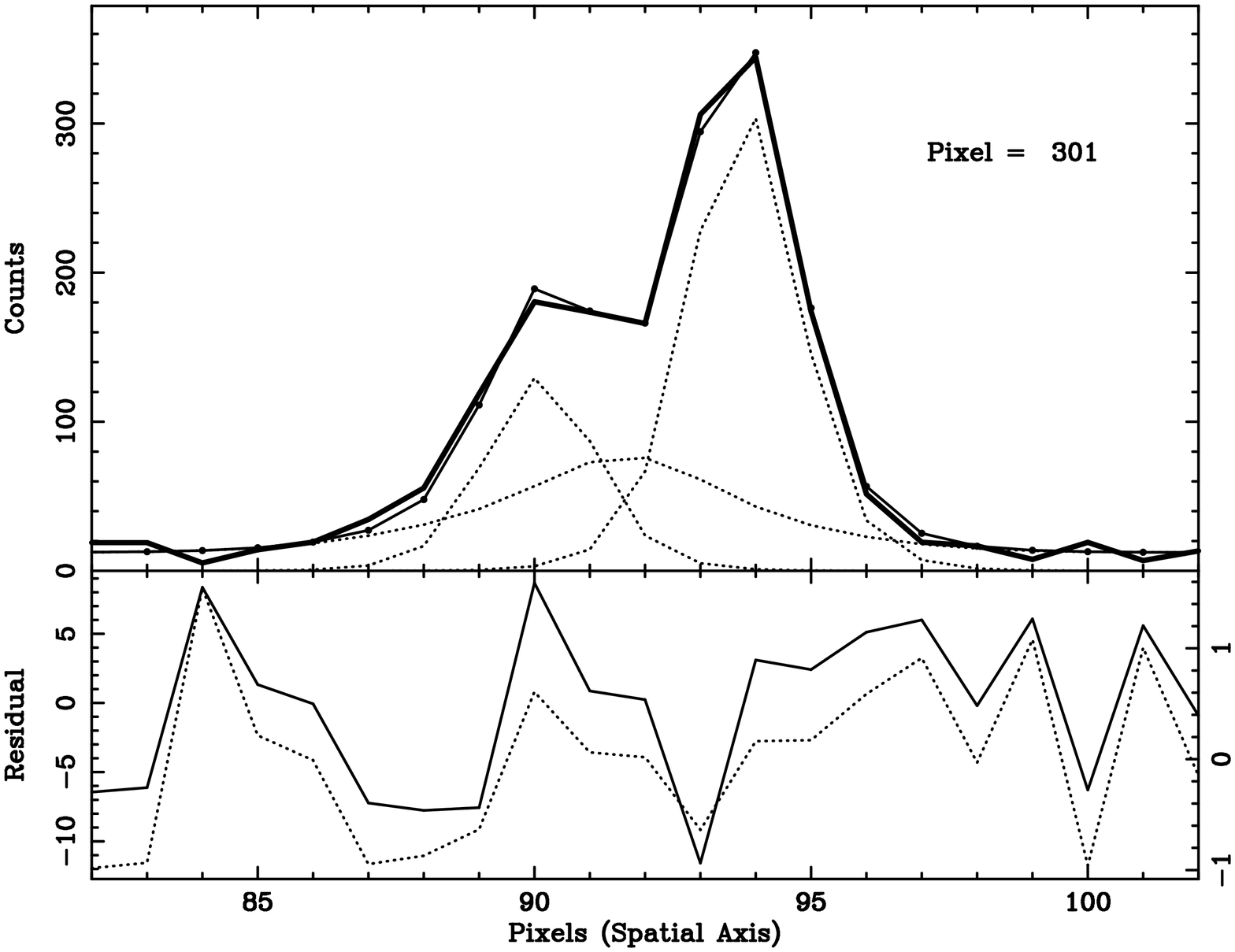,width=3.5in,angle=270}
}
\caption[Model--fit to a wavelength section]{ 
Upper panel, model--fit (solid line) including contributions from the
quasar and galaxy components (dashed lines) and the data (heavy solid
line). $\chi^2(\lambda_j)=0.545$. In the wings, at Pixels$<86$ and
Pixels$>98$ only the galaxy contributes to the model fit. The lower
panel presents the residuals (data-model), in counts (solid line and
left hand axis), and normalized by the noise (dashed line and right
hand axis).}
\label{profile_slice}
\end{figure}

\subsection{Initial Reductions}

Bias subtraction and flat-fielding were accomplished using standard
procedures within the {\tt IRAF} reduction package. To accomplish sky
subtraction, the sky level was estimated directly from the CCD frames
by defining strips either side of the the quasar spectra. These strip
were 15 pixels in extent, starting at least 30 pixels from the quasar
spectra. At this distance the contribution of the extended galaxy is
negligible.  The average of the median values of the sky in each strip
was adopted as the sky value at each wavelength increment. There was no
evidence of a gradient in the sky along the slit. For each
object--frame a companion error--frame, recording the noise per pixel
\---\ including contributions from sky and object counts, CCD
read--noise and flat-fielding uncertainties \---\ was retained.

Wavelength and flux calibration proved straightforward.  Uncertainties
in the wavelength calibration, derived from the residuals of
individual arc lines about the cubic--spline fits employed were
0.2\AA~in the blue and 0.05\AA~in the red. Relative fluxes, to an
accuracy of $\sim$10\% across the spectra, were obtained following
repeat observations of standard stars.  Given the narrow slits used
and the systematic loss of light due to atmospheric dispersion it was
decided that the final spectra should not be flux calibrated.

Following sky subtraction, each CCD frame consisted of two quasar
spectra superimposed upon the more extended lensing galaxy.
Coefficients to perform wavelength and flux conversions were retained
but not applied to the target frames, \ie~no rebinning or scaling that
would compromise the independence of the noise from pixel to pixel was
performed. The proximity of the quasar images to one--another, $\sim
1\scnd2$, and the presence of the galaxy light precluded the use of
conventional extraction procedures. Instead, a procedure whereby the
composite profile was modeled by three components, the two unresolved
quasars and the galaxy, was employed.

\subsection{Spectroscopic Extraction}\label{reduction_method}

\begin{figure*}
\centerline{
\psfig{figure=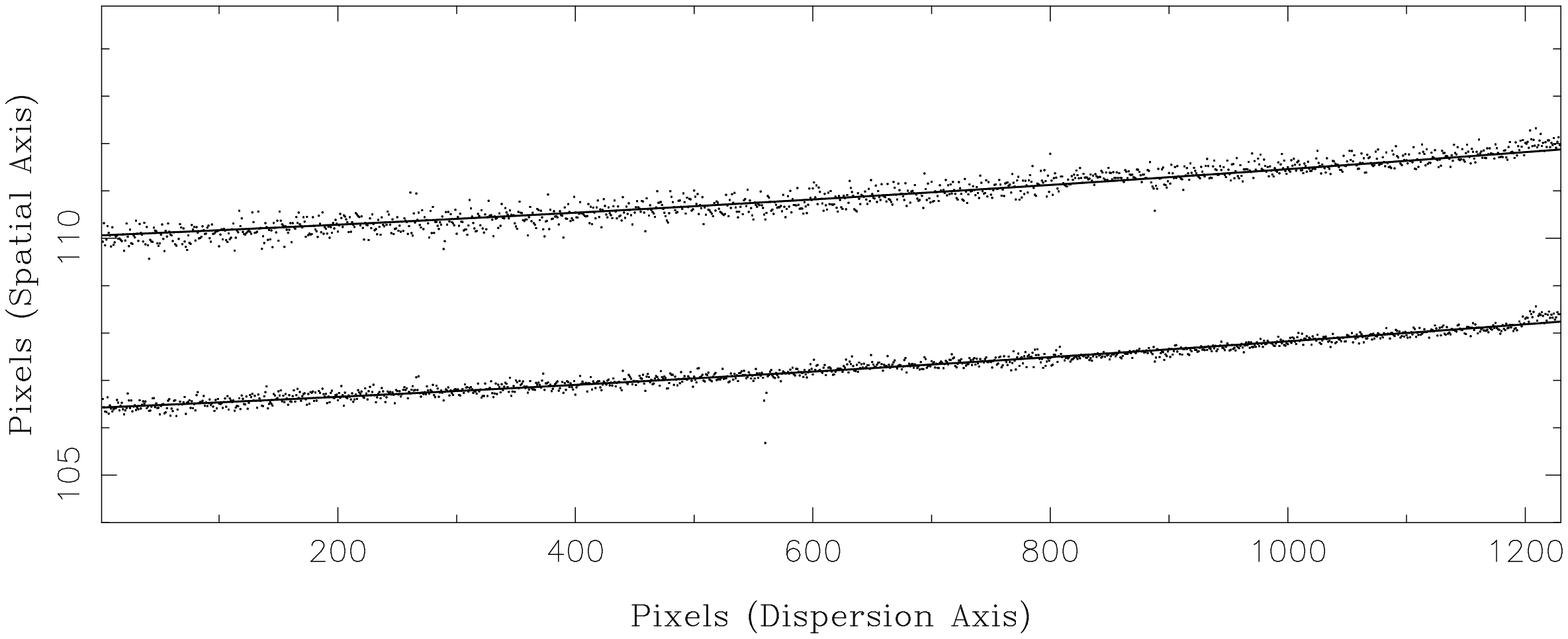,width=5.5in,angle=270}
}
\caption[Spectral image centroids]{ 
Image centroids determined from the first stage of the fitting procedure 
and the analytic third--order polynomial representations which allow the
centroids to be specified at each wavelength increment. The data is from
the red spectrum of images A+C, taken in 1994.}
\label{qso:centroid}
\end{figure*}

The key to the effective application of a multi--component model
fitting procedure to estimate the contributions of the two quasar
components and the galaxy at each wavelength increment is to reduce the
number of free parameters to a minimum. In principle, the number of
free parameters is large as three object profiles, their locations
along the slit and their relative heights all need to be determined.
However, a significant reduction can be achieved by taking advantage of
the well--behaved and slowly varying properties of the spectra as a
function of wavelength.

Specifically, the centroids of the three components along the slit
(pixel position on the Y--axis of the CCD frame) varies smoothly (see
below) as a function of wavelength (pixel position on the X--axis of
the CCD frame). An analytic fit to the location of the centroids of the
components as a function of wavelength allows the three component
centroid positions to be specified prior to the model fitting
at each wavelength.

The profiles of the two (unresolved) quasars components are identical,
reducing the number of independent profiles to be determined from
three to two.  Furthermore, the variation in profile--shape with
wavelength, arising from the combined effects of the
wavelength--dependent atmospheric seeing and scattering in the
spectrograph, is well--behaved, allowing the shape of the quasar
profiles to be specified at each wavelength increment.

Thus, implementation of the model--fitting procedure involves two
stages: In the first, the systematic variation of the positions and
shape of the component profiles is determined and the shape of the
component profiles estimated. In the second stage the contributions of
the three components, essentially the relative intensity of the three
component profiles, at each wavelength (pixel) increment is calculated
using a non--linear minimization routine.

In the first stage of the model--fitting procedure, an unweighted fit,
neglecting any galaxy component, is used to establish the centroids of
the point spread function of the quasar components. The location of the
image centroids can then be represented by an analytic function,
reducing the number of free parameters to three: the width of the
seeing profile, and the normalization of each of the quasar components.
The concluding stage of the procedure consists of a fully weighted
model--fit, including a model for the galaxy contribution.  Before
considering the details of the quasar and galaxy profiles adopted, an
overview of the entire procedure is provided.  The analysis
proceeds as follows:

\begin{enumerate}

\item An unweighted fit of a model consisting of two identical image
profiles (the quasars) is undertaken for each wavelength increment in
the CCD frame. Five free parameters, the profile centroids, the profile
heights and the profile width (effective seeing) are determined via a
Fortran90 implementation of a modified version of the {\tt AMEOBA}
algorithm of Press~\etal~(1988).  The initial location for the profile
centroid (along the slit) is specified interactively for the first
wavelength increment. For subsequent wavelengths, the location of the
centroid from the previous wavelength increment is employed. The
contribution from the galaxy profile is ignored at this stage. The
unweighted fit is very sensitive to the peaks of the (quasar) point
spread functions.

\item An analytic function is then fit to the resulting profile
centroid positions with the constraint that the profile separation
remains constant (independent of wavelength). A third order polynomial
representation produces excellent results (Figure~\ref{qso:centroid}).
The zeroth--order terms were allowed to differ, while higher--order
terms were forced to match for each of the quasars.

\item A model fit, as in step (i) above, is then performed, but now
variance--weighting is employed and the two image centroids derived
from the polynomial representation determined in step (ii).  Three free
parameters are calculated, the profile width and the heights of the two
quasar profiles.

\item An analytic function is then fit to the resulting profile width
as a function of wavelength. As for the centroid positions the
dependence of the profile width on wavelength is smooth and well
behaved allowing a very accurate representation using a simple analytic
function.

\item Finally, a variance--weighted model fit is employed with the
image centroids and profile width determined from their respective
analytic representations. There are only three parameters remaining,
the heights of the two quasar profiles and the galaxy profile.  The
relative contributions of each component from the fit averaged over
one CCD exposure are illustrated in Figure~\ref{profile_slice}.

An accurate representation of the quasar and galaxy profiles and a
quantitative measure of goodness--of--fit between data and model are
required for the model--fitting approach to work effectively. 

\end{enumerate}

\subsection{Model Components}

\subsubsection{The Quasar Point Spread Functions}\label{psf}

The very high signal--to--noise ratio spectra of the standard stars,
obtained using the identical spectrograph configuration on the same
night as the observations of Q2237+0305, were used to define the form
of the point spread function (PSF) for unresolved point sources. An
analytic function, $p(r)$, consisting of a Gaussian core with
exponential wings described the point spread function to very high
accuracy. $p(r)$ is defined:

\begin{equation} 
p(r)=
\left \{ \begin{array}{ll} K e^{ - \frac{1}{2} \left(\frac{ r }
{ \sigma}\right)^2} & {r <
r_{lim}} \\ P e^{ - \frac{ r } { \delta } } & {r > r_{lim}}
\end{array}
\right. , 
\end{equation} 
where $r$ is the distance from the centre of the profile and $r_{lim}$
is the distance at which the two profiles join~\cite{irwin1985}.  By
ensuring that the profile is continuous in the zeroth and first
derivatives at $r_{lim}$ then
\begin{equation} 
P = \frac{ K }{N} e^{2 \log{N}},\quad \delta = \frac{
\sigma }{ \sqrt{ 2 \log{N} } },
\label{seeing_function} 
\end{equation} 
where $N$ is the fraction of the psf peak at which the profiles join.
The condition that the profile is continuous in both the zeroth and
first derivatives reduces the number of independent parameters to
four; the central position, $x_c$; radial scale length, $\sigma$;
normalization, $K$; and the join fraction, $N$.

Variations in atmospheric seeing produce changes in the PSF that are
essentially confined to the central Gaussian core and the PSFs of all
three standard star spectra could be modeled by fixing the join
fraction, $N$ equal to $35\%$ of the peak profile intensity, thereby
reducing the number of independent parameters necessary to describe
the profile to three.

As described in the previous section, the centroid and width (the
$\sigma$ parameter) of the quasar profiles as a function of wavelength
are determined prior to the final stage of the model fitting. Thus the
profile height, or normalization, is the sole parameter to be
determined for each of the two quasar spectra.

\subsubsection{The Galaxy Profile}\label{galaxy_cmpt}

\begin{figure}
\centerline{
\psfig{figure=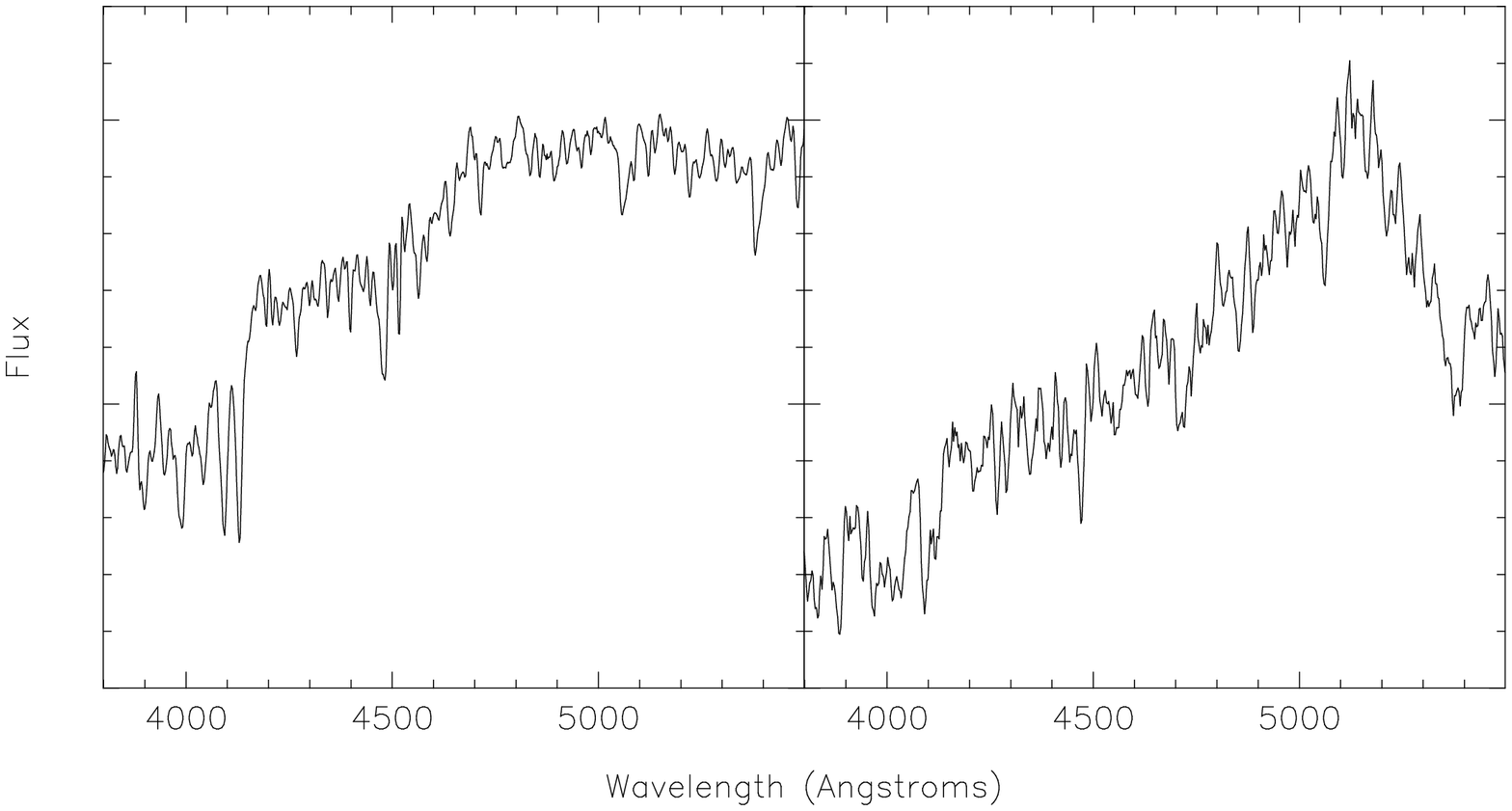,width=3.5in,angle=270}
}
\caption[Galaxy Spectrum]{ 
The left panel shows a spectrum of NGC 3379, an E0 galaxy from the
Kennicutt~(1992) spectral atlas, redshifted to $z = 0.039$, the
redshift of the lensing galaxy in the 2237+0305 system.  The right
panel presents a galaxy spectrum extracted from the 1991 data for the
B+D images. The spectrum has not been flux calibrated. A number of
absorption features are common to both spectra, \eg Ca{\footnotesize
II} H{\footnotesize\&}K~$\lambda\lambda3968,3934$, the $4000{\rm \AA}$
break, the G-band $\lambda 4300$, H$\delta$ $\lambda4102$, H$\beta$
$\lambda 4861$, and Mg{\footnotesize I}~$\lambda\lambda5173,5175$.}
\label{qso:kenni}
\end{figure}

The galaxy profile was modeled using a de Vaucoulers (1953) ${\rm
r^{\frac{1}{4}}}$ profile.  The core radius was chosen to $r_e =
7\scnd0$, and the ellipticity to $\epsilon = 0.31$, in accordance with
the observation of Racine~(1991). The profile was smoothed with a PSF
of the form determined from the observations of the standard stars
[Section~\ref{psf}] with the profile width defined by the value
calculated from the quasar spectra in the CCD frames.  Virtual slits
of $0\scnd7$ and $0\scnd91$ were defined across the resulting surface
brightness distribution at the same orientation as the actual
observations. The light distribution was integrated across the slit at
high resolution along the slit to give the spatial profile of galaxy
light observed through the slit. The profile could be well reproduced
using an analytical model consisting of two Gaussian components and a
constant DC level, and this analytic representation was employed in
the model--fitting procedure, with the height of the galaxy model
profile as the free parameter. The normalization of the galaxy
component and the two quasar components are obtained simultaneously in
the model--fitting procedure, thereby producing a galaxy spectrum in
addition to the two quasar spectra.  The galaxy spectrum extracted
from one of the data frames, uncorrected for instrumental response, is
presented in the right hand panel of Figure~\ref{qso:kenni}. For
comparison, the spectrum of NGC 3379~\cite{kennicutt1992}, an E0
galaxy, redshifted to $(z = 0.039)$ is shown in the left panel of the
figure. The number of absorption features in common between the
spectra, including Ca{\footnotesize II}
H{\footnotesize\&}K~$\lambda\lambda3968,3934$, the $4000{\rm \AA}$
break, the G-band $\lambda4300$, H$\delta$ $\lambda4102$, H$\beta$
$\lambda 4861$, and Mg{\footnotesize I}~$\lambda\lambda5173,5175$,
indicates the success of the fitting procedure in recovering a
spectrum with the absorption line properties expected to be present in
the galaxy core.

\subsection{Goodness--of--Fit}\label{gof}

A standard $\chi^2(\lambda_j)$ statistic was employed to ascertain the
goodness--of--fit of the model to the data. The analysis procedure
involves one--dimensional model fits, $M(\lambda_j,y_i)$, to the observed
distribution counts, $S(\lambda_j,y_i)$, in the spatial direction at
fixed wavelength, (\ie~index $\lambda_j$ is fixed).

\begin{eqnarray}
\chi^2(\lambda_j) = \frac{1}{N_{pix}}\sum_{i=1}^{N_{pix}} \
\frac{[ S(\lambda_j,y_i)-M(\lambda_j,y_i) ]^2}{\sigma^2(\lambda_j,y_i)} ,
\label{merit_function}
\end{eqnarray}
where the model, $M(\lambda_j,y_i)$ is:
\begin{equation}
M(\lambda_j,y_i) = \int^{\scriptscriptstyle y_i+\frac{\Delta
y}{2}}_{\scriptscriptstyle y_i-\frac{\Delta y}{2}} [
Q_1(\lambda_j,y')+Q_2(\lambda_j,y')+G(\lambda_j,y') ] dy',
\label{spectral_modelling}
\end{equation}
where $Q_j$ and $G$ are analytic models for the two quasar components
and the galaxy respectively. These functions are integrated numerically
over a pixel, where $\Delta y$ is the pixel width. The noise, $\sigma^2(\lambda_j,y_i)$, is:

\begin{equation}
\sigma^2(\lambda_j,y_i) = S(\lambda_j,y_i ) + Sky(
\lambda_j ) + R^2,
\label{noise_est}
\end{equation}
where the first term represents the $\sqrt{N}$ noise from the quasar
and galaxy components, the second component the $\sqrt{N}$ noise from
the sky, and the third component the read noise, $R$, from the CCD.

The integrator used throughout was a Fortran90 implementation of the
routine {\tt DQAG}, a Gauss-Konrod general-purpose, globally adaptive
integrator, available from the numerical mathematical software database
GAMS~\footnote{\tt http://gams.nist.gov/}.  This integrator proved very
capable in standard integral tests, especially with the polynomial type
integrands utilized in the evaluation of the $\chi^2$ statistic.

\begin{figure*}
\centerline{
\psfig{figure=fig_a4.ps,width=5.5in,angle=270}
}
\caption[Fits to fake data]{Results of applying the analysis to the
synthetic quasar data. Lower panel shows the determinations of the
quasar image centroids. Middle panel presents the quasar profile width
measures.  Top panel shows the quasar spectra recovered. The solid
line in each panel represents the input values.}

\label{fitted_vals}
\end{figure*}

\begin{figure}
\centerline{
\psfig{figure=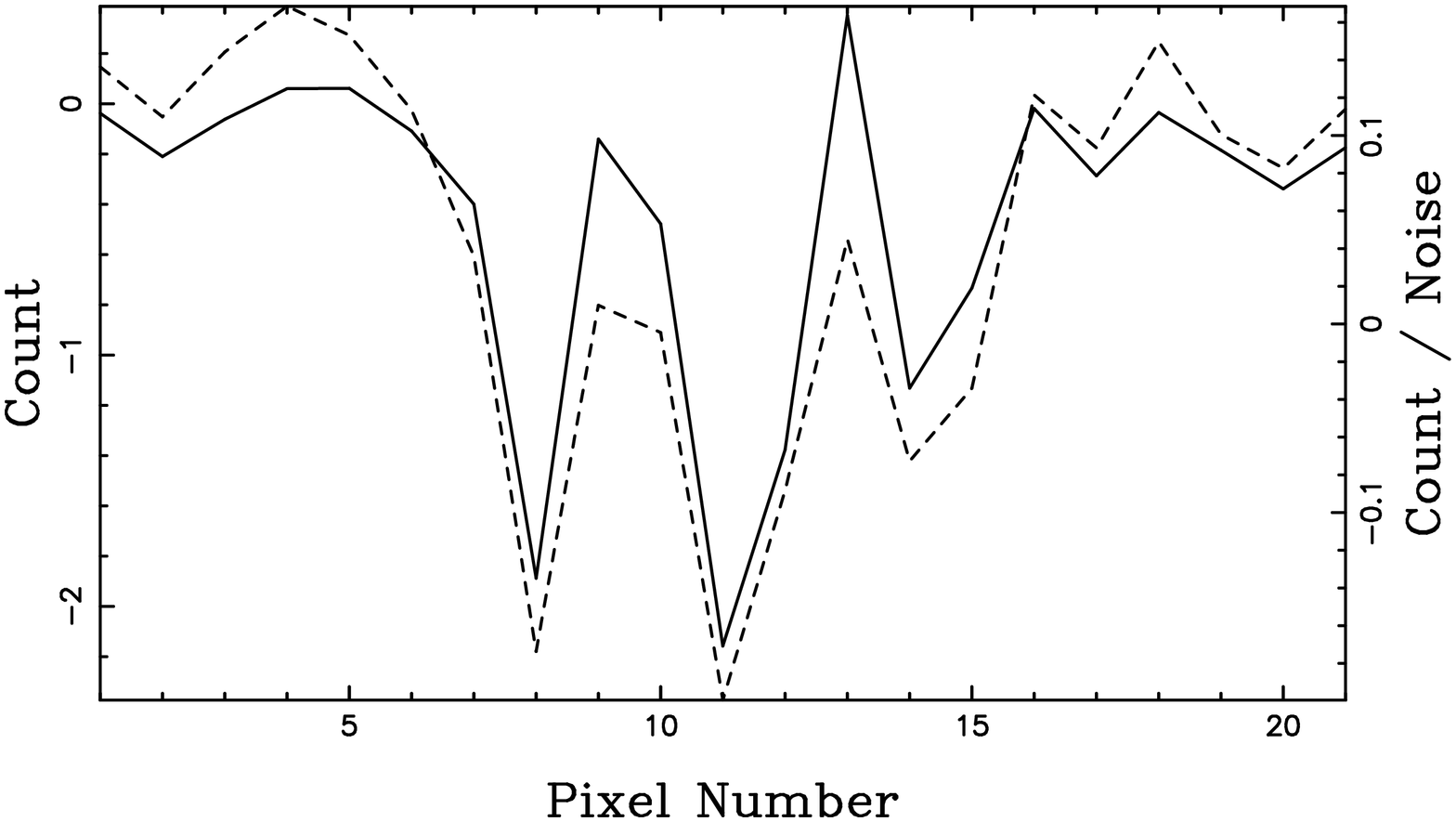,width=3.5in,angle=270}
}
\caption[Fitting Residuals]{ 
The summed residuals from the model--fit to the synthetic data.
Residuals (data-fit) in counts, in the spatial direction, averaged
over the dispersion axis are shown as the solid line (left hand
axis). The residuals, normalised by the noise at each pixel are shown
as the dotted line (right hand axis).}
\label{residual}
\end{figure}

\subsection{Fitting Procedure Applied to Synthetic Data}

To verify the effectiveness of the extraction technique, simulated
observations, closely mimicking the properties of the actual data, were
generated and the ability of the fitting procedure to recover the
(known) input parameters tested. The two quasar components, utilizing
the composite quasar spectra derived by Francis (1991), were
represented by PSFs with peak heights of 60 and 120 counts, and
profile width $\sigma = 1$ pixel. The components were separated by 3.7
pixels, with one centroid at pixel position 98.15, while the second
was placed at 101.85. The galaxy component had a peak height of 15
counts and the spatial profile of Section~\ref{galaxy_cmpt}. Poisson
noise from the object, a sky noise of 6 counts, combined with a read
noise $\sigma_n = 10$ counts, were included. This resulted in a
signal--to--noise ratio of $\sim$2.5--3.5 for the simulated quasar
components, a value lower than the actual observations.

The results of the fitting procedure are illustrated in
Figure~\ref{fitted_vals}. The lower panel presents the measured image
centroids from the first stage of the fitting procedure. ${\rm P_1}$
and ${\rm P_2}$, are the mean positions and the $1\sigma$ scatter
about the mean. The profile centroid separation is $3.69\pm0.22$
pixels. The middle panel shows the recovery of the profile seeing
width, $\sigma$. The top panel presents the individual spectra
recovered by the analysis with the actual input spectra shown for
comparison.  Figure~\ref{residual} presents the corresponding mean
residual (solid line and left--hand axis) in the spatial
direction. The dashed line represents the mean residual count
normalized by the noise (right--hand axis) at each position. The
residuals are small, typically $\la15\%$, compared to the signal in
the spectrum. The median value of the chi-squared merit function
(Equation~\ref{merit_function}) over the fit is 1.20, a reasonable
value considering the piece--wise and non--optimal nature of the
fitting procedure.

Recovery of the two quasar components from an artificial data set with
a signal--to--noise ratio and level of galaxy contamination closely
matching that of the actual data gives confidence in the spectra
obtained by applying the procedure to the observational data.

\subsection{Robustness}

Differences will occur between the models employed in the spectral
fitting, namely the point spread function and the surface brightness
distribution of the galaxy, and the true functions.  Such differences
can introduce systematic errors into the extracted quasar spectra,
affecting quantities such as the emission line equivalent widths.
However, the parameters employed in the modeling are derived from
observations of the lensing system and standard stars, and such
systematic errors should be small. These errors should, therefore, in
no--way effect the overall conclusions of this paper.

\vfil

\end{document}